\documentclass[twocolumn,showpacs,preprintnumbers,amsmath,amssymb,prb]{revtex4}

\usepackage{graphicx}
\usepackage{dcolumn}
\usepackage{bm}

\begin{document}

\preprint{APS/123-QED}

\title{Fermi liquid behavior and weakly anisotropic superconductivity in
electron-doped cuprate Sr$_{1-x}$La$_{x}$CuO$_{2}$
}

\author{Kohki H. Satoh$^{1}$}
 \email{ksatoh@post.kek.jp}
\author{Soshi Takeshita$^{2}$}%
\author{Akihiro Koda$^{1,2}$}%
\author{Ryosuke Kadono$^{1,2}$}%
\author{Kenji Ishida$^{3}$}%
\author{Sunsen Pyon$^{4}$}%
\author{Takao Sasagawa$^{4}$}%
\altaffiliation[Present affiliation: ]{Materials and Structures
Laboratory, Tokyo Institute of Technology}
\author{Hidenori Takagi$^{4}$}%
\affiliation{%
$^{1}$Department of Materials Structure Science,
The Graduate University for Advanced Studies, Tsukuba, Ibaraki 305-0801, Japan\\
$^{2}$Institute of Materials Structure Science, High Energy Accelerator Research
Organization, Tsukuba, Ibaraki 305-0801, Japan\\
$^{3}$Department of Physics, Graduate School of Science, Kyoto University,
Kyoto 606-8502, Japan\\
$^{4}$Department of Advanced Materials Science, University of Tokyo, Kashiwa,
Chiba 277-8561, Japan
}%

\date{\today}

\begin{abstract}
The microscopic details of flux line lattice state studied by muon spin rotation 
is reported in an electron-doped high-$T_{\rm c}$ cuprate superconductor, 
Sr$_{1-x}$La$_{x}$CuO$_{2}$ (SLCO, $x=0.10$--0.15).
A clear sign of phase separation between magnetic and  non-magnetic phases is 
observed, where the effective magnetic penetration depth [$\lambda\equiv\lambda(T,H)$] 
is determined selectively for the latter phase.  
The extremely small value of $\lambda(0,0)$ 
and corresponding large superfluid density ($n_s \propto \lambda^{-2}$) 
is consistent with presence of a large Fermi surface with carrier density of 
$1+x$, which suggests the breakdown of the ``doped Mott insulator" 
even at the ``optimal doping" in SLCO.  Moreover, a relatively weak anisotropy 
in the superconducting order parameter is suggested by the field dependence
of $\lambda(0,H)$. These observations strongly suggest that the superconductivity 
in SLCO is of a different class from hole-doped cuprates.
\end{abstract}


\maketitle

The question whether or not the mechanism of superconductivity 
in electron-doped ($n$-type) cuprates is common to that 
in hole-doped ($p$-type) cuprates 
is one of the most interesting issues
in the field of cuprate superconductors, which is yet to be answered.
This ``electron-hole symmetry'' has been addressed by many experiments
and theories since the discovery of $n$-type cuprate superconductors.\cite{asy}
In the theoretical models assuming strong electronic correlation where 
the infinitely large on-site Coulomb interaction ($U\rightarrow\infty$)
leads to the Mott insulating phase for the half filled band, 
the correlation among the doped carriers is projected into the 
$t$-$J$ model in which the mechanism of superconductivity does not 
depends on the sign of charge carriers.\cite{tjmodel,zgrc}  This is in marked
contrast to the models starting from Fermi liquid (= normal metal) state,
where such symmetry is irrelevant to their basic framework.\cite{frmlqd}
Experimentally, recent advent in crystal growth techniques 
and that in experimental methods
for evaluating their electronic properties triggered 
detailed measurements on $n$-type cuprates,
reporting interesting results suggesting certain 
differences from $p$-type ones,
such as the observation of a commensurate spin fluctuations in neutron 
scattering study or the nonmonotonic $d$-wave superconducting order parameter in
ARPES measurement.\cite{neutron,ARPES}

The effective magnetic penetration depth ($\lambda$) is one of the most important 
physical quantities directly related with the superfluid density ($n_{\rm s}$),
\begin{equation}
\frac{1}{\lambda^{2}} = \frac{n_{\rm s}e^{2}}{m^{*}c}, \label{lmdns}
\end{equation} 
which is reflected in the microscopic 
field profile of the flux line lattice (FLL) state in type II superconductors. 
Considering that the response of $n_s$ against various perturbations
strongly depends on the characters of the Cooper pairing, 
the comparison of $n_s$ between two types of carriers 
might serve as a testing ground for the electron-hole symmetry.
However, the study of FLL state in $n$-type cuprates 
such as T'-phase $RE_{2}$CuO$_{4}$ compounds ($RE$= Nd, Pr, Sm, etc.),
is far behind that in $p$-type cuprates 
because of strong random local fields from rare-earth ions 
which mask information of CuO$_{2}$ planes regarding both superconductivity
and magnetism against magnetic probes such as muon.
In this regard, infinite-layer structured Sr$_{1-x}$La$_{x}$CuO$_{2}$ (SLCO) 
is a suitable compound for detailed 
muon spin relaxation and rotation ($\mu$SR) study of electron-doped
systems, as it is free from magnetic rare-earth ions. 

A recent $\mu$SR study on SLCO with $x=0.10$ ($T_{\rm c}\simeq40$~K) reported 
a relatively large $n_s\propto\lambda_{ab}^{-2}$
[$\lambda_{ab}(T\rightarrow0)\sim$116~nm] 
as compared 
to $p$-type cuprates,\cite{PSI} strongly suggesting that $n$-type cuprates
belong to a different class in view of the $n_{\rm s}$ 
versus $T_{\rm c}$ relation.\cite{Uemura:91}
On the other hand, another $\mu$SR study showed appearance of a spin glass-like 
magnetism over a wide temperature range including superconducting phase,\cite{Kojima}
which might have also affected the result of Ref.~\onlinecite{PSI}.
In this report, we demonstrate by $\mu$SR measurements under both zero and 
high transverse field that SLCO exhibits a phase separation into magnetic 
and non-magnetic phases, where the superconductivity occurs predominantly in the latter.
Our measurement made it feasible to evaluate $\lambda$ reliably as it was
selectively determined for the non-magnetic phase of SLCO.

Meanwhile, the paring symmetry of order parameter,
which is one of the most important issues in discussing the electron-hole symmetry,
still remains controversial in $n$-type cuprates.
A number of groups reported $s$-wave symmetry in SLCO,\cite{STM,SHeat}
which is in marked contrast to the $d_{x^2-y^2}$ symmetry well established in 
$p$-type cuprates. The pairing symmetry can be examined 
by measuring the temperature/field variation of $\lambda$($T,H$) 
as an {\sl effective} value observed by $\mu$SR:
it reflects the change of $n_{\rm s}\equiv n_{\rm s}(T,H)$ due to quasiparticle 
excitation and/or nonlocal
effect associated with anisotropic order parameter.\cite{Kadono:07,Kadono:07_2}
Here, we show evidence that the order parameter
in SLCO is not described by simple isotropic $s$-wave pairing nor 
that of pure $d_{x^2-y^2}$.

\begin{figure}
\includegraphics[width=\linewidth]{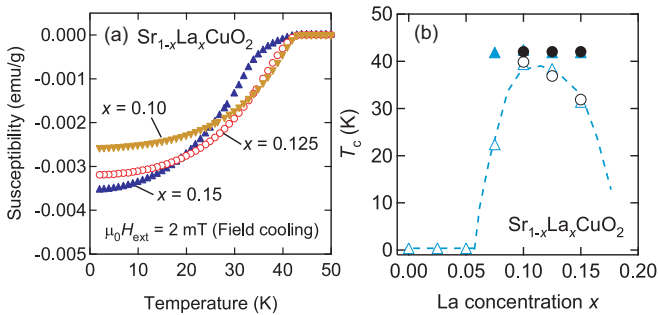}
\caption{\label{fig:MTandTc} (Color online)
(a) Magnetic susceptibility of SLCO with  
$x$=0.10, 0.125 and 0.15 under 20~G. 
(b) La concentration dependence of $T_{\rm c}$. 
Closed symbols show $T_{\rm c~onset}$, and open symbols
show $T_{\rm c~bulk}$, respectively. 
An earlier result\cite{kawashima} (triangles) is also 
quoted for comparison. Dashed line is guide to eyes.}
\end{figure}

Powder samples of SLCO ($x$ = 0.10, 0.125, and 0.15) were prepared by high pressure 
synthesis under 6~GPa, 1000~$^\circ$C. 
They were confirmed to be of single phase by powder X-ray diffraction, where 
a small amount of LaCuO$_{2.5}$ phase (LCO2.5, less than a few \%) was 
identified. The length of $a$ and $c$ axes 
showed almost linear change with $x$, indicating successful substitution of Sr with 
La for carrier doping.\cite{GEr}  As displayed in Fig.~\ref{fig:MTandTc}(a), 
the susceptibility ($\chi_0$) measured by SQUID magnetometer implies that 
the onset of superconductivity is nearly 42~K and 
least dependent on $x$, whereas the bulk $T_{\rm c}$
determined by the maximum of $d\chi_0/dT$ 
varies with $x$ [see Fig.~\ref{fig:MTandTc}(b)], 
which reproduces earlier results.\cite{kawashima,karimoto}  The $x$ dependence of
bulk $T_{\rm c}$ suggests that the sample is close to the optimal doping for $x=0.1$.

The $\mu$SR experiment was performed 
on the M15 beamline at TRIUMF (Vancouver, Canada),
where measurements under zero and longitudinal field (ZF and LF)
were made to investigate magnetic ground state of SLCO.  
Subsequently, those 
under a high transverse field (HTF, up to 6~T) were made to study the
FLL state in detail.  
In ZF and LF measurements, a pair of scintillation counters (in backward and forward geometry relative to the initial muon polarization that was parallel to the beam direction) were employed
for the detection of positron emitted preferentially to the 
muon polarization upon its decay.
In HTF measurements, sample was at the center of four position counters placed around the beam axis, 
and initial muon spin polarization was perpendicular to the muon 
beam direction so that the magnetic field can be applied along the beam direction without interfering with beam trajectory.  A veto counter system was employed to eliminate background 
signals from the muons that missed the sample, which was crucial 
for samples available only in small quantities 
such as those obtained by high-pressure synthesis. 
For the measurements under a transverse field, the sample was 
field-cooled to the target temperature 
to minimize the effect of flux pinning.

\begin{figure}
\includegraphics[width=1.0\linewidth]{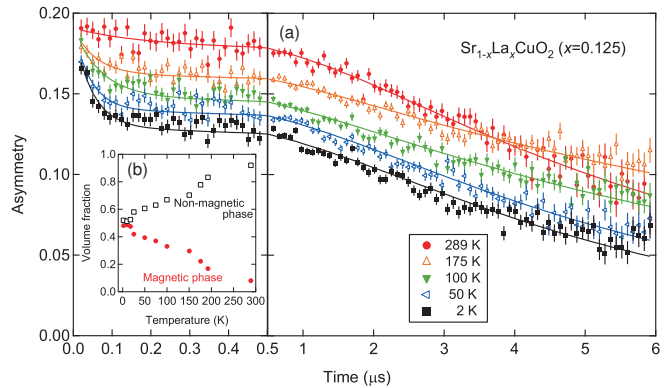}
\caption{(Color online)
(a) ZF-$\mu$SR spectra in the sample with $x$=0.125.
Inset (b) is temperature dependence of the volume fraction of 
magnetic and non-magnetic phase.}
\label{fig:ZFmSR} 
\end{figure}

\begin{figure}
\includegraphics[width=1.0\linewidth]{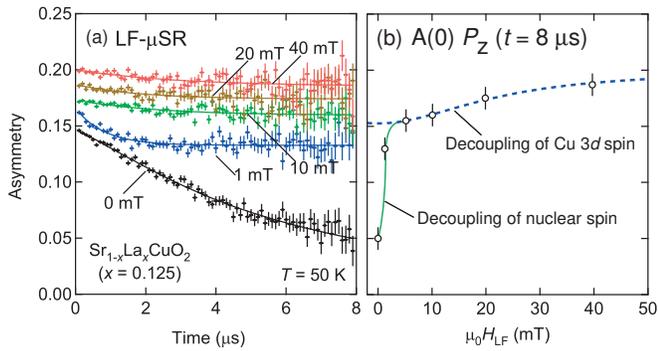}
\caption{\label{fig:LFmSR} (Color online)
(a) LF-$\mu$SR spectra in the sample with $x$=0.125.
(b) Two kinds of decoupling behavior at 50~K. The dashed line
is fitting curb by Eq.~(\ref{LFfit})}
\end{figure}

Fig.~\ref{fig:ZFmSR}(a) shows ZF-$\mu$SR spectra 
for the sample with $x$=0.125,
where no spontaneous muon precession is observed as sample is
cooled down to 2~K.  Instead, fast muon spin depolarization can be 
identified between 0$<t<$0.2~$\mu$s, which develops with decreasing
temperature.
LF-$\mu$SR spectra in Fig.~\ref{fig:LFmSR}(a) shows that the depolarization is 
quenched in two steps as a function of field strength, at first near 
a few mT due to nuclear magnetic moments  and 
secondly around 10$^1$ mT.
The asymptotic behavior of $P_{\rm z}(t)$ under random local fields 
$H$ (with an isotropic mean square, $\langle H_{\rm x}^2 \rangle=\langle H_{\rm y}^2 \rangle=\langle H_{\rm z}^2 \rangle=\frac{1}{3}\langle H^2 \rangle$) as a function of external magnetic field $H_{\rm LF}$ is approximately given by the follows equation,
\begin{equation}
P_{\rm z}(t\rightarrow \infty)\approx \frac{H_{\rm LF}^2+\langle H_{\rm z}^2 \rangle}
{H_{\rm LF}^2+\langle H^2 \rangle}
=\frac{H_{\rm LF}^2+\frac{1}{3}\langle H^2 \rangle}{H_{\rm LF}^2+\langle H^2 \rangle},
\label{LFfit}
\end{equation} 
and we estimated the magnitude of $ \sqrt{\langle H^2 \rangle} \equiv \overline{H}_{\rm int}$ from the behavior of $P_{\rm z}(t\rightarrow \infty)$ as 39(3)~mT [the best fit with Eq.~(\ref{LFfit}) is shown in Fig.~\ref{fig:LFmSR}(b)]. 
This is consistent with the fast initial depolarization rate estimated by $\gamma_\mu\overline{H}_{\rm int}=33(3)$ MHz (where  $\gamma_\mu=2\pi\times135.53$ MHz/T is the muon gyromagnetic ratio).  The origin of $\overline{H}_{\rm int}$ can be uniquely attributed to the 
localized moments Cu atoms, 
where the effective moment size is
0.15(1)~$\mu_{\rm B}$.  The almost negligible depolarization for the asymptotic component implies that spin fluctuation rate is much smaller than $\gamma_\mu\overline{H}_{\rm int} $ at 50~K. Thus, ZF/LF-$\mu$SR results strongly suggest that 
the sample that exhibits superconductivity  
has also static magnetic phase.
The magnetic region enlarges to a halfway partition at low temperature
(as seen in Fig.~\ref{fig:ZFmSR}(b)).  
We note that a common tendency was observed for $x$=0.10 and 0.15.

\begin{figure}
\includegraphics[width=0.90\linewidth]{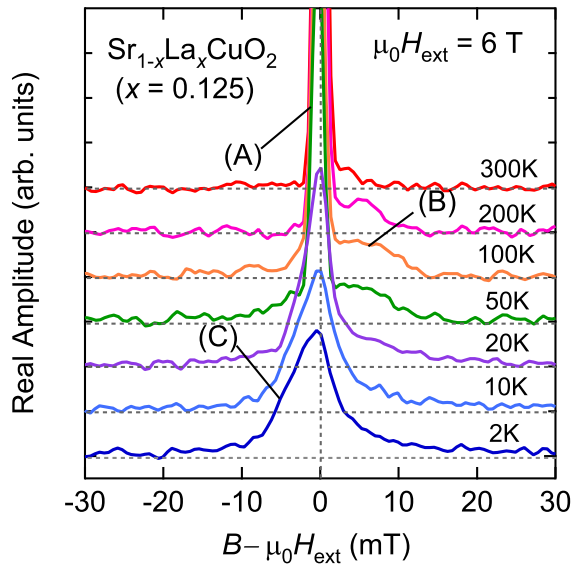}
\caption{\label{fig:fft6t} (Color online) Fast Fourier Transform of 
HTF-$\mu$SR spectra under 6~T [after filtering the
artifacts due to a finite time window for transform ($0\le t\le 6$ $\mu$s)].
The peaks labeled A, B and C correspond to non-magnetic/non-superconducting,
 magnetic, and FLL phases, respectively.}
\end{figure}

In HTF-$\mu$SR, 
each pair of counters (right-left, upward-downward)
observes  time-dependent muon spin polarization, $\hat{P}(t)$,
projected to $x$ or $y$ axis perpendicular to the beam direction
(with a relative phase shift of $\pi/2$).  Inhomogeneity  of 
magnetic field distribution $B({\bf r})$ leads to depolarization 
due to the loss of phase coherence among muons probing different parts
of $B({\bf r})$.  Using a
complex notation,  $\hat{P}(t)$ is directly provided using the 
spectral density distribution for the internal field, $n(B)$, 
\begin{eqnarray}
\hat{P}(t) = P_{x}(t)+iP_{y}(t) = \int^{\infty}_{-\infty} n(B)
e^{i(\gamma_{\mu}Bt-\phi)}dB,
\end{eqnarray}
where $n(B)$ is defined as a spatial average ($\langle \rangle_{r}$) of the delta function,
\begin{eqnarray}
n(B)=\langle \delta(B-B({\bf r}))\rangle_{r},
\end{eqnarray}
and $\phi$ is the initial phase of muon spin rotation.
Then, the real part of fast Fourier transform (FFT) of $\mu$SR time
spectrum corresponds to $n(B)$, namely,
\begin{eqnarray}
n(B)=\Re\int^{\infty}_{-\infty} \hat{P}(t)e^{-i(\gamma_{\mu}Bt-\phi)}dt.
\end{eqnarray}
Fig.~\ref{fig:fft6t} shows the real amplitudes obtained by the FFT 
of HTF-$\mu$SR spectra, which contain information on $n(B)$.
The narrow central peak (labeled A) is the signal 
from muons stopped in a non-magnetic (and/or non-superconducting) phase where 
the frequency is equal to that of 
the external field ($\mu_{0} H_{\rm ext}=6$ T) with a linewidth 
determined by random nuclear dipolar fields besides the effect
of limited time window ($0\le t\le6$ $\mu$s).
A broad satellite peak 
(labeled B) appears on the positive side of the central peak, when temperature
is lowered below 300 K.  This corresponds to the fast depolarization in time domain.
The ZF/LF-$\mu$SR spectra in Figs.~\ref{fig:ZFmSR},\ref{fig:LFmSR}  demonstrates that this 
satellite comes from a magnetic phase in which quasistatic
random magnetism of Cu electron spins develops.

While the FFT spectra were useful to examine the overall feature 
of $n(B$),  the actual data analysis was carried out in time domain
using the $\chi^2$-minimizing method. 
As inferred by Fig.~\ref{fig:fft6t}, the $\mu$SR spectra in the normal state can be reproduced 
by a sum of two Gaussian dumping signals,
\begin{eqnarray}
\hat{P}_{\rm n}(t) &=& \sum^2_{k=1}f_{k}\int^{\infty}_{-\infty} n_k(B)e^{i(\gamma_{\mu}Bt-\phi)}dB\nonumber\\
&=& \sum^2_{k=1}f_{k}\exp(-\sigma_{k}^2 t^2/2)
e^{i(\omega_{k}t-\phi)},\label{twoGs}
\end{eqnarray}
where $f_{k}$ is the relative yield proportional to the fractional volume of each phase, 
$\sigma_{k}$ is the linewidth, and 
$\omega_{k}$=$\gamma_{\mu}\overline{B}_k$ with $\overline{B}_k$  being the mean value of 
local magnetic field following a Gaussian distribution, 
\begin{equation}
n_k(B) = (\sqrt{2\pi}\sigma_k)^{-1}\exp[-\gamma_\mu^2(B-\overline{B}_k)^2/2\sigma_k^2].\nonumber
\end{equation}
It is inferred from the $\chi^2$-minimizing fit of the time spectra by Eq.~(\ref{twoGs})
that the volume fraction of magnetic phase increases 
toward low temperature monotonously in place of non-magnetic phase
and becomes nearly a half at 50~K. 
This is clearly not due to the LCO2.5 impurity phase, 
considering the small volume fraction of LCO2.5 and its known N$\acute{\rm e}$el 
temperature ($\sim$125~K).\cite{Ladder}
The magnetic volume fraction is independent of $H_{\rm ext}$ 
at 50~K where the sample is in the normal state.
Thus, the appearance of the satellite peak demonstrates
the occurrence of a phase separation into magnetic 
and non-magnetic domains in the normal state of SLCO.

Taking the result in the normal state into consideration, 
we analyzed the $\mu$SR spectra in the superconducting phase.
In the FLL state of type II superconductors, one can reasonably 
assume that muon stops randomly over the length scale of vortex lattice, 
and serves to provide a random sampling of inhomogeneity due to FLL formation. 
In the modified London (m-London) model, $B({\bf r})$ 
is approximated as a sum of magnetic inductions from isolated vortices,

\begin{eqnarray*}
B_{\rm v}({\bf r})=B_{0}\sum_ {\bf K}\frac{e^{-i {\bf K} \cdot {\bf r}}}{1+K^2\lambda^2}F(K,\xi_{v})
\end{eqnarray*}
where ${\bf K}$ are the vortex reciprocal lattice vectors,
$B_{0}$  ($\simeq H_{\rm ext}$) is
the average internal field, $\lambda\equiv\lambda(T,H)$ is the {\sl effective} London 
penetration depth depending on temperature and field, 
and $F(K,\xi_{\rm v})=\exp(-K^2\xi_{\rm v}^2/2)$ 
is a nonlocal correction term with $\xi_{\rm v}$ ($\simeq\xi$)
being the cutoff parameter for the magnetic field distribution;
the Gaussian cutoff generally provides satisfactory agreement with data.
The density distribution $n(B)$ in this case 
is characterized by the Van Hove singularity originating from the 
saddle points of $B_{\rm v}({\bf r})$ with a negative shift primarily determined 
by $\lambda$, and that corresponds to the peak (seen as a shoulder) 
labeled C in Fig.~\ref{fig:fft6t}.
Thus, the signal from the FLL state can be readily separated from 
other phases at large $H_{\rm ext}$ as they exhibit different frequency 
shifts with each other. The FFT spectra below $T_{\rm c}$ also 
indicate that the domain size of the superconducting
phase is much greater than that determined by $\lambda$.

It is known that the m-London model is virtually identical to 
the Ginzburg-Landau (GL) model for large $\kappa=\lambda/\xi$ ($\xi$ is
the GL coherence length) and at low magnetic fields ($H_{\rm ext}/H_{c2}<0.25$, 
with $H_{c2}$ being the upper critical field).\cite{Brandt,SonierRev}
Meanwhile, according to a reported value of the upper critical field for SLCO
($\mu_{0}H_{\rm c2}$=12 T, in Ref.\onlinecite{Hc2}), 
the field range of the present measurements ($0\le \mu_0 H_{\rm ext}\le 6$ T) 
might exceed the above mentioned boundary, and thus the use of
the GL model would be more appropriate.  However, the m-London model has 
certain advantages over the GL model in practical application to the analysis: 
for example, we can avoid further complexity of analysis due to introduction of the field-dependent {\it effective} coherence length.\cite{Kadono:07}
We also stress that the discrepancy in the analysis results has been  
studied in detail between these two models, and now it is well established  
that m-London model exhibits a systematic tendency of slight overestimation of 
$\lambda$ at higher fields due to a known cause.\cite{Kadono:07,SonierRev}
The discussion on the present result will be made below considering 
this tendency.  

Another uncertainty comes from the fact that the FLL symmetry 
in SLCO is not known at this stage, and 
it might even depend on the magnitude of external field as has been found
in some other cuprates.\cite{Brown,Gilardi}
However, since we do not observe any abrupt change of lineshape 
nor the increase of $\chi^2$ in the fits  (irrespective of model)  
associated with the alteration of FLL symmetry with varying field,\cite{Nb3Sn, YNBC} 
we can reasonably assume that the FLL symmetry remains the same  
throughout entire field range.  Moreover, the observed lineshape 
is perfectly in line with the hexagonal FLL, without showing any 
sign of squared FLL (e.g., a large spectral weight at the lower field side 
of the central peak in the absence of nonlocal effect\cite{Aegerter:98}, or 
an enhanced weight at the central peak 
associated with the strong nonlocal effect\cite{YNBC}).  Therefore, the FLL
symmetry has been assumed to be hexagonal in the following analysis.

The $\mu$SR spectra in the FLL state were analyzed by fit analysis using
%
\begin{eqnarray}
\hat{P}(t)& = &  \hat{P}_{\rm v}(t)+\hat{P}_{\rm n}(t),\\
 \hat{P}_{\rm v}(t)& \equiv &f_{\rm v}e^{-\sigma_{p}^2 t^2}
\int{n_{\rm v}(B)e^{i(\gamma_{\mu}Bt-\phi)}}dB,
\end{eqnarray}
\begin{equation}
n_{\rm v}(B)=\langle \delta(B-B_{\rm v}({\bf r}))\rangle_{r},
\end{equation}
where $f_{\rm v}$ 
is the volume fraction of FLL phase, $\sigma_{\rm p}$ represents 
the contribution from the distortion of FLL due to vortex pinning and that
due to nuclear random local fields, and 
$\hat{P}_{\rm n}(t)$ is that defined in Eq.~(\ref{twoGs}). 
The parameters including $f_{\rm v}$, $\lambda$, $\xi_{\rm v}$, $\sigma_p$, 
$f_k$, $\sigma_k$ and $\omega_k$ were determined by the $\chi^2$-minimization
method with good fits as inferred from the value of reduced $\chi^2$
close to unity.  (More specifically, in order to reduce the uncertainty 
for the analysis of data below $T_c$,  $\omega_k$ was fixed to the value 
determined by the data above $T_c$.)  The magnitude of line broadening due to
vortex pinning ($\sigma_p$) was relatively small 
(typically 30--40\% of the frequency shift for the shoulder C in Fig.~\ref{fig:fft6t}). 
This was partly due to relatively short $\lambda$ and associated large asymmetry 
in $n(B)$, and thereby 
the correlation between these parameters turned out to be small
except at lower fields ($\mu_0 H_{\rm ext}\le1 $ T) where the 
spectra exhibit stronger relaxation due to greater linewidth of  
$n(B)$ and stronger vortex pining (leading to larger $\sigma_p$).

\begin{figure}
\includegraphics[width=\linewidth]{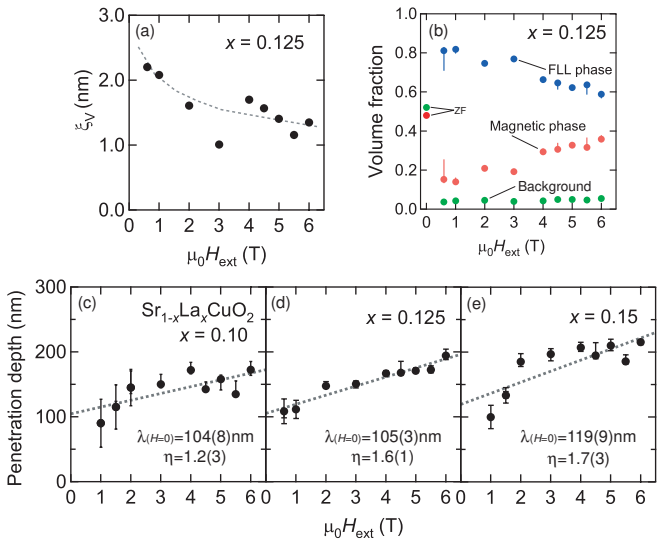}
\caption{\label{fig:lmd}  (Color online)
Field dependence of cutoff parameter (a)
and volume fraction of each phase (b) at 2~K for the sample with $x$=0.125, where 
dashed curves are guide to eyes.
(c)--(e): Field dependence of effective 
penetration depth at 2~K for $x$=0.10, 0.125 and 0.15, respectively.
Dashed lines are a linear fit (see text).}
\end{figure}

Fig.~\ref{fig:lmd}(a) shows a decreasing tendency of $\xi_{\rm v}$ with
increasing field, which is understood as a shrinkage of vortex core due to 
vortex-vortex interaction.\cite{SonierRev}
Fig.~\ref{fig:lmd}(b) shows the field
dependence of fractional yield for each phase at 2~K.  
With increasing field, the FLL phase appears to be transformed into the 
magnetic phase. However, it must be noted that there is a discontinuous
change between ZF ($\sim50$\%) and HTF-$\mu$SR ($\sim60$--80\%).
Since no field dependence is observed  for the volume fraction in the 
normal state ($T\sim$50~K), the reduction of the magnetic fraction at lower
fields is attributed to the overlap of magnetic domains with vortex cores: 
the magnetic domains would serve as pinning centers for vortices more
effectively at lower fields due to the softness of FLL.
The increase of magnetic fraction with increasing field is then readily
understood as a result of decreasing probability for vortices to overlap with random 
magnetic domains at higher fields, because the relative density of vortices as well as the 
rigidity of FLL would increase. This also suggests that the mean domain size of the
magnetic phase is considerably smaller than the FLL spacing ($=69$ nm at 0.5 T).

\begin{figure}
\includegraphics[width=\linewidth]{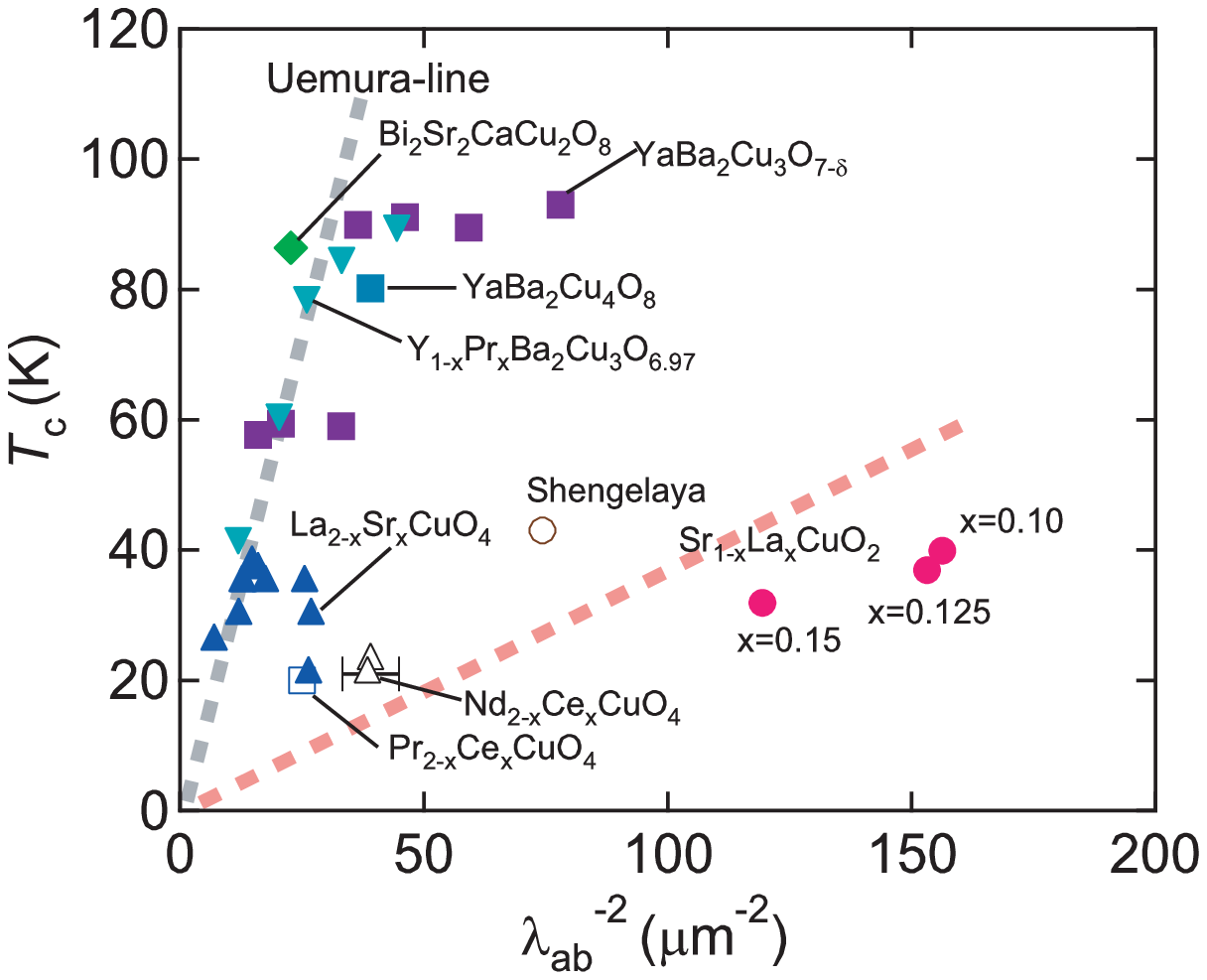}
\caption{\label{fig:Uemuraplot} (Color online) 
$T_{\rm c}$ vs $\lambda^{-2}$ for various cuprate superconductors.
Closed circles represent our result, 
whereas open circle is that of Ref.~\onlinecite{PSI}.
Open squares and triangles are for other
$n$-type cuprates\cite{PCCO_Homes,NCCO_Nugroho,NCCO_Homes},
closed upward triangles for La$_{2-x}$Sr$_{x}$CuO$_{4}$ 
(Ref.~\onlinecite{LS15_Luke,LS15_Aeppli,LS_Schneider, LS_Pana})
and downward ones for Y$_{1-x}$Pr$_{x}$Ba$_{2}$Cu$_{3}$O$_{6.97}$ 
(Ref.~\onlinecite{Seaman_YPBCO}).
Square symbols for YBa$_2$Cu$_3$O$_{y}$ and diamond 
for Bi$_{2}$Sr$_{2}$CaCu$_{2}$O$_{8}$  
(Ref.~\onlinecite{YBCO_Hardy, YBCO_Sonier, Uemura_YBCO, Keller_YBCO}).}
\end{figure}

Fig.~\ref{fig:lmd} (c)--(e)
show the field dependence of $\lambda$ in each compounds.
While the London penetration depth is a physical constant uniquely 
determined by local electromagnetic response, 
$\lambda$ in our definition [Eq.~(\ref{lmdns})] is a variable parameter,
as $n_{\rm s}$ depends on temperature ($T$) and external magnetic field ($H$).
Therefore, we introduce an effective penetration depth,
$\lambda(T,H)$ with an explicit reference to $T$ and $H$ dependence. 
It is clear in Fig.~\ref{fig:lmd} (c)--(e) that $\lambda(H)=\lambda(2\: {\rm K},H)$ tends to increase with increasing external field.
Here, one may further notice a tendency that $\lambda(H)$ increases
more steeply below $\sim$2 T in the case of $x=0.10$ and $0.15$.
However, these points at lower fields are also associated with larger error bars 
probably because of the stronger depolarization in the time domain.
The value extrapolated to $\mu_{\rm 0}H_{\rm ext}=0$ [$\lambda(0)$]
is estimated by a linear fit with a proper consideration of the
uncertainty associated with these errors, and the result is 
indicated in Fig.~\ref{fig:lmd}.
These values (104--119~nm) turn out to be significantly shorter than 
the earlier result\cite{PSI}
(hereafter, the inplane penetration depth $\lambda_{\rm ab}$ is approximated by
an equation $\lambda \simeq 1.3\lambda_{ab}$, according to Ref.~\onlinecite{lmdab}). 
In qualitative sense, however, our result supports the earlier suggestion 
of a large discrepancy for SLCO from the quasi-linear relation 
between $T_{\rm c}$ and $n_{\rm s}$ 
observed over a wide variety of $p$-type cuprates.\cite{Uemura:91}
The anomaly becomes more evident when they are mapped to the 
$T_{\rm c}$ vs $\lambda^{-2}$ plot, 
as shown in Fig.~\ref{fig:Uemuraplot}.
They are far off the line followed by the data of $p$-type cuprates,
suggesting that $n$-type SLCO belongs to a class of superconductors different from 
that of $p$-type cuprates.   

It is well established that the carrier concentration, $p$, of $p$-type cuprates 
nearly corresponds to that of the doping value $x$ while 
$x \le$ 0.20.\cite{Fukuzumi}
In contrast, a recent ARPES measurement on an $n$-type cuprate,
Nd$_{2-x}$Ce$_{x}$CuO$_{4\pm \delta}$, have revealed that  
small electron pockets
($p \simeq x$) observed for $x$=0.04 sample is replaced  by a 
large Fermi surface (corresponding to $p\simeq 1+x$)
for $x$=0.10 and 0.15 samples.\cite{Armitage}
When $m^{*}$ is assumed to be comparable with 
that of $p$-type cuprate ($m^{*}\simeq3m_{\rm e}$),
$n_{\rm s}$ can be estimated using Eq.~(\ref{lmdns}), yielding
$1.3 \times 10^{22}{\rm cm}^{-3}$ in SLCO with $x$=0.125 [where $\lambda(0)$
is determined with the best accuracy].
This corresponds to $p\ge 0.70$, and   
an order of magnitude larger than that of $p$-type cuprates.\cite{LS15_Aeppli,SonierRev} A better correspondence to $p\simeq1+x=1.125$ would be 
attained when $m^{*}\simeq4.8m_{\rm e}$.
Thus, the present result is yet another evidence for a large
Fermi surface in SLCO.  This is also in line with some 
recent experimental results for $n$-type superconductors.
For example, resistivity ($\rho$) in the normal state
shows a Fermi liquid-like temperature dependence ($\rho \propto T^{\rm 2}$)
common to ordinary metals,\cite{rhoT} and a metallic Korringa law 
has been revealed by NMR study under high magnetic fields.\cite{Korringa}
These observations coherently suggest that the $n$-type cuprates  
cannot be regarded as the doped Mott-insulators, but they might be better  
understood as in the normal Fermi liquid state
already at the optimal doping ($x \sim$0.1).

The increase of $\lambda$ with increasing external field is a 
clear sign that the superconducting order parameter is not 
described by that of simple isotropic $s$-wave paring for single-band 
electrons.\cite{Kadono:07}
One of the possible origins for the field dependent $\lambda$ is the presence 
of nodal structure in the order parameter [$|\Delta({\bf k})|=0$ 
at particular ${\bf k}$] 
that leads to the field-induced 
quasiparticle excitation due to the quasiclassical Doppler shift.\cite{Volovik}
The quasiparticle energy spectrum is shifted by the flow of supercurrent
around vortex cores to an extent $\delta E=m {\bf v}_{\rm F} \cdot  {\bf v}_{\rm s}$,
where ${\bf v}_{\rm F}$ and ${\bf v}_{\rm s}$ are the 
Fermi velocity and superfluid velocity, respectively. This gives rise to the 
pair breaking for $|\Delta({\bf k})|<\delta E$ and associated reduction
of $n_{\rm s}$.
The presence of nodes 
also leads to a nonlocal effect in which $\lambda$ is affected by the 
modification of supercurrent near the nodes where the coherence 
length $\xi_0({\bf k})=\hbar v_{\rm F}/\pi\Delta({\bf k})$ exceeds the 
local London penetration depth.\cite{Amin:98}
For the comparison of magnitude for the field-induced effect, we use
a dimensionless parameter $\eta$ deduced by fitting data in Fig.~\ref{fig:lmd} using 
$\lambda(h)= \lambda(0)[1+\eta h]$ with $h = H/H_{\rm c2}$.
Provided that $\eta$ is dominated by the presence of gap nodes,
the magnitude of $\eta$ at lower fields is roughly proportional to the 
phase volume of the Fermi surface where $|\Delta({\bf k})|<\delta E$.
As seen in Fig~\ref{fig:lmd}, 
$\eta$ in SLCO is definitely greater than zero irrespective
of $x$, taking values between 1.2--1.7.
It is noticeable that these values are considerably smaller than $\eta \simeq 6$ ($\mu_{0}
H_{\rm ext}< 2~T$) observed in YBa$_{2}$Cu$_{3}$O$_{6.95}$ (YBCO) that has a typical 
$d_{x^2-y^2}$-wave gap symmetry. The situation remains true even when 
one considers i) the nonlocal effect that tend to reduce $\eta$ at high magnetic fields
($\eta \simeq 2$ for $\mu_{0}H_{\rm ext}> 2$~T),\cite{nonlocal} and ii) a possible
overestimation of $\lambda$ at higher fields due to the extended use of the m-London model that also leads to the overestimation of $\eta$ [e.g., $\eta$ based on the m-London model is greater than that on the GL model by 0.23(7) in NbSe$_2$ (Ref.~\onlinecite{SonierRev}), and 0.6(2) in YB$_6$ (Ref.~\onlinecite{Kadono:07})].

Interestingly, the relatively small value of $\eta$ is in line with the recent 
suggestion by ARPES measurement
on another $n$-type superconductor, Pr$_{0.89}$LaCe$_{0.11}$CuO$_{4}$ (PLCCO), that 
the order parameter $\Delta(k,\psi)$ has a steeper
gradient at the nodes along azimuthal ($\psi$) direction than that for
the $d_{x^2-y^2}$ symmetry.\cite{ARPES}
Since the phase volume satisfying $|\Delta({\bf k})|<\delta E$ is
inversely proportional to $d|\Delta(k,\psi)|/d\psi$ at the node, we have
\begin{equation}
\eta\propto\left( \frac{d|\Delta(k,\psi)|}{d\psi}\right)^{-1}_{\psi(|\Delta|=0)}.
\end{equation}
Assuming a situation similar to PLCCO and that $\eta$ observed in YBCO represents
a typical value for $d_{x^2-y^2}$-wave gap, 
our result suggests that the gradient $d|\Delta(k,\psi)|/d\psi$ in SLCO is 1.2(3)--5.0(3)
times greater than that at the node of $d_{x^2-y^2}$-wave gap.  
However, it is clear that further
assessment by other techniques that are more sensitive to the symmetry of the
order parameters are necessary to discuss the details of gap structure in SLCO.

In conclusion, it has been revealed by the present $\mu$SR study that a phase separation occurs 
in an electron-doped cuprate superconductor, Sr$_{1-x}$La$_{x}$CuO$_{2}$ 
($x$=0.10, 0.125 and 0.15), where nearly half of the sample volume
exhibits magnetism having no long-range correlation while the rest remains
non-magnetic.  The superconductivity occurs predominantly in the non-magnetic domain, 
where the effective magnetic penetration depth evaluated by using a modified-London 
model is much shorter than that of other $p$-type cuprates.  This suggests a large
carrier density corresponding to $1+x$ and accordingly the breakdown of the 
Mott insulating phase in SLCO and other $n$-type cuprates even at their optimal doping. 
The field dependence of $\lambda$ suggests that the superconductivity of SLCO 
is not described by single-band $s$-wave pairing.  The magnitude of the dimensionless
parameter, $\eta$ ($\propto d\lambda/dH$), is qualitatively in line  
with nonmonotonic $d$-wave superconducting gap  
observed in other $n$-type cuprates.

We would like to thank the staff of TRIUMF for technical support
during the $\mu$SR experiment and Takano-group of ICR-Kyoto Univ.
(M. Takano, Y. Shimakawa, M. Azuma, I. Yamada and K. Oka)
for useful advice concerning sample preparation.
This work was partially supported by the Grant-in-Aid for 
Creative Scientific Research and 
the Grant-in-Aid for Scientific Research on 
Priority Areas by the Ministry of Education, 
Culture, Sports, Science and Technology, Japan.

\appendix

\newpage 

\end{document}